\documentclass[conference]{IEEEtran}
\usepackage{pbox}

\usepackage{cite}

\ifCLASSINFOpdf
  \usepackage[pdftex]{graphicx}
\else
\fi
\usepackage{amsmath}
\usepackage{algorithm}
\usepackage{algorithmic}
\makeatletter
\def\BState{\State\hskip-\ALG@thistlm}
\makeatother

\usepackage{array}
\usepackage{url}

\usepackage{xcolor}
\begin{document}
%
\title{Real-Time Electric Vehicle Smart Charging at Workplaces: A Real-World Case Study}



%
\author{\IEEEauthorblockN{Nathaniel Tucker\IEEEauthorrefmark{2},
Gustavo Cezar\IEEEauthorrefmark{3}, and
Mahnoosh Alizadeh\IEEEauthorrefmark{2}}
\IEEEauthorblockA{\IEEEauthorrefmark{2}Department of Electrical and Computer Engineering, 
University of California, Santa Barbara,
California, 93106, USA}
\IEEEauthorblockA{\IEEEauthorrefmark{3}SLAC National Accelerator Laboratory, GISMo Group, California, 94025, USA}
}


\maketitle


\begin{abstract}
We study a real-time smart charging algorithm for electric vehicles (EVs) at a workplace parking lot in order to minimize electricity cost from time-of-use electricity rates and demand charges while ensuring that the owners of the EVs receive adequate levels of charge. Notably, due to real-world constraints, our algorithm is agnostic to both the state-of-charge and the departure time of the EVs  and uses  scenario generation to account for each EV's unknown future departure time as well as certainty equivalent control to account for the unknown EV arrivals in the future. Real-world charging data from a Google campus in California allows us to build realistic models of charging demand for each day of the week. We then compare various results from our smart charging algorithm to the status quo  for a two week period at a Google parking location.
\end{abstract}


%
\IEEEpeerreviewmaketitle
\makeatletter
\def\blfootnote{\xdef\@thefnmark{}\@footnotetext}
\makeatother

\section{Introduction}
\label{section: introduction}

Many large companies across the United States are installing Electric Vehicle Supply Equipment (EVSEs) within their parking lots to allow for employees and visitors to charge their Electric Vehicles (EVs) during their stay \cite{workplace_EV}. There are numerous benefits that can come from workplace charging stemming from the highly flexible nature of EV charging at workplaces \cite{8635950, 9281711} (i.e., EVs are plugged in for long durations and their charging can be shifted depending on other conditions including behind-the-meter solar generation or electricity rates). For example, Google (who has installed EVSEs at several of its Bay Area locations for employee EV charging) allows employees to leave their EVs connected to the EVSE for long durations and does not force them to unplug the EV after reaching full charge, whereas other workplaces might require employees to move EVs after charging is complete.  The approach of not enforcing EVs to be removed once fully charged may require more investment in EVSEs to satisfy the needs of the employees, but the investment would allow for greater flexibility that can be harnessed by a smart charging algorithm to benefit the workplace (by providing financial savings) and the utility (by enabling its customers to respond to price signals) \cite{ferguson2018optimal, tucker2019online}. 

The Grid Integration, Systems, and Mobility (GISMo) group within the SLAC National Accelerator Laboratory and the Smart Infrastrucuture Systems Lab (SISL) at UCSB are working on a project  to study the impacts of electric vehicles on the distribution system and designing tools and algorithms to be used by workplace charging station operators to efficiently manage asset operations and minimize electricity charges. 
In collaboration with Google, the project has access to historical charging data from previous years at multiple Bay Area workplace parking locations for Google (within PG\&E's service territory). The locations exhibit typical workplace charging behavior in that most session start times occur around 8 to 9 A.M. and most session end times occur between 4 and 6 P.M. The dataset contains 15-minute interval data for over 10,000 charging sessions and between 700 and 1000 sessions per month. The datasets include pertinent information such as charging session start times, end times, 15 minute average power delivered, total energy delivered, and many others. 



\textit{Challenges:} There are several key challenges to designing an algorithm to schedule the charging profiles of numerous EVs. First and foremost, the algorithm must run in real-time without knowledge of the future EV arrivals. The algorithm must adapt its planned power schedules as more information is revealed (i.e., as more EVs arrive to the parking lot). Second, contrary to most smart charging algorithms presented in the literature, the algorithm must be able to function with limited information from each EV \cite{7586042}, \cite{6919255}. Specifically, when an EV plugs in, the algorithm does not get access to the EV's State of Charge (SoC) nor does it know the EV's future departure time (most level 2 chargers do not sense EV SoC nor do they request user input for future departure times). As such, our smart charging algorithm must predict how much maximum charge an EV may consume as well as the EV's future departure time. Such challenges have been acknowledged in past papers including \cite{8585856, 8586075, 8585744, 8586132}. Third, all of the EV charging schedules within a parking lot are coupled due to the local transformer capacity constraint \cite{POWELL2020115352}. As such, the algorithm cannot over-allocate power at any given time; therefore, the algorithm should make use of a model of the future EV arrivals to avoid over allocating power due to unexpected arrivals.

\textit{Contribution:} In this paper, we present an EV smart charging algorithm  for workplace parking lots equipped with EVSEs that operates in real-time to minimize electricity cost from time-of-use electricity rates and demand charges while ensuring that the owners of the EVs receive adequate levels of charge and the entire system safely operates within the local transformer capacity constraint. Notably, our algorithm is SoC and departure time agnostic and uses both scenario generation to account for each EV's unknown future departure time as well as certainty equivalent control to account for the unknown EV arrivals in the future. We build models from the Google dataset for each day of the week and our algorithm uses these models as the expected future when optimizing the EV charging schedules. 


\section{Problem Description}
\label{section: problem description}



\subsection{Base Optimization}
\label{sec:base}

Before we discuss the smart charging algorithm that operates in real-time with a predicted model of the future, let us consider the simpler problem of scheduling the charging power to a single EV if we know the arrival time, departure time, and maximum energy. Specifically, let us break one day into 15 minute segments (96 segments total) and let us denote $t=1,\dots,T$ as the time steps each day and $T=96$. 
Furthermore, let $x$ be a Tx1 indicator vector where the $t$-th element is 1 if the EV is plugged in at time $t$ and 0 if not. Additionally, let $p$ be a Tx1 vector where the $t$-th element corresponds to the electricity rate at time $t$ from the local grid (\$/kWh).  Last, let us denote our decision variable $e$ as a Tx1 vector where the $t$-th element indicates how much energy will be delivered to the EV in time epoch $t$.  With  this notation, we can formulate the charge scheduling optimization problem for the singular EV as follows:

\begin{subequations}
\begin{align}
    &\hspace{2ex}\max_{e}  && \hspace{-2ex} w_1 \log (e^Tx+1) - w_2 p^T e 
    \label{eqn:obj}
    \\
    & \hspace{2ex}\nonumber\textrm{subject to:}
    \\
    & && \hspace{-2ex} 0 \leq e \leq e_{max},
    \label{eqn:0}
    \\
    & && \hspace{-2ex} e_{del\_min} \leq e^Tx \leq e_{del\_max}.
    \label{eqn:deliveredmin}
\end{align}
\end{subequations}

The objective function \eqref{eqn:obj} has two weight coefficients, $w_1$ and $w_2$, for each of the two terms. The first term corresponds to the utility that the owner of the EV receives for the energy that their EV receives. The logarithmic utility term was chosen to model the diminishing returns in user utility for EVs receiving excessive amounts of energy (e.g., the first 20kWh charged is more valuable to the EV owner than the second 20kWh). The second term corresponds to the cost of the energy that is purchased from the local grid (we will add demand charges in the next section). The weights, $w_1$ and $w_2$, let us adjust the relative importance of user utility from receiving energy and energy cost. Constraint \eqref{eqn:0} ensures that the energy delivered at each time step is non-negative and not greater than the energy $e_{max}$ that the EVSE can deliver in a single timestep. Constraint \eqref{eqn:deliveredmin} ensures that the amount of energy the EV receives is greater than a minimum amount $e_{del\_min}$ and less than a maximum amount $e_{del\_max}$.


\section{Real-Time Smart Charging Algorithm}
\label{section: algorithm}

In this section, we modify the base optimization presented in Section \ref{section: problem description} that will be solved at each 15 minute time step $t=1,\dots,96$ each day. Specifically, assume that at an arbitrary time $t$, there are currenly $I$ EVs plugged in at the location.

\textit{Departure time scenario generation:} When an EV plugs in to an EVSE, we generate $N$ potential departure times for that EV and create a scenario in our optimization problem for each potential departure time and solve the optimization across all scenarios. As time progresses, if a potential departure time is no longer feasible (i.e., the potential departure time is the current time step and the EV has not yet departed), then that scenario is removed from the optimization via dynamic scenario weights (let $C_n$ be a weight coefficient for each scenario that is set to 0 if the scenario is no longer feasible.). Furthermore let $x_{i,n}$ be the Tx1 binary vector that indicates when EV $i$ is available to charge in scenario $n$.  

\textit{Certainty equivalent control for future EV arrivals:} We make use of our dataset to generate a model for an average day that consists of estimated arrival times, departure times, and energy requests for each day of the week. We then use these daily models in the real-time optimization to account for the unknown future EV arrivals. Specifically, at time $t$, let us assume that there are $J$ EVs in the certainty equivalent daily model that are expected to arrive in the future. Let $x_j$ be the Tx1 binary vector indicating when EV $j$ is available to charge. The decision variables that determine how much energy is delivered at a given time $t$ are Tx1 vectors labeled as $e_1,\dots,e_I$ for the actual EVs plugged in and $e_{I+1},\dots,e_{I+J}$ for the future EV arrivals from the model. Additionally, let $p_d$ be the demand charge (\$/kW) to be assessed on the monthly peak load. The optimization at time $t$ can be written as:

\begin{subequations}
\begin{align}
    \label{eqn:obj mpc}\max_{\substack{e_1,\dots,e_I\\e_{I+1},\dots,e_{I+J}}}
    &\sum_{i=1}^I \hspace{-1pt}\sum_{n=1}^N \frac{1}{C_n} \Big[w_1 \log (e_i^{\;T} x_{i,n}+1) - w_2 p^T e_i\Big] \\
    +\nonumber&\hspace{-1ex}\sum_{j=I+1}^{I+J} \Big[w_1 \log (e_j^{\;T} x_{j}+1) - w_2 p^T e_j\Big] - w_2 p_d \hat{e}_{inc}
    \\
    \nonumber\textrm{subject to:}
    \\
    &0 \leq e_k \leq e_{max}, \hspace{19.5pt} \forall k=1,\dots,I+J
    \label{eqn:0 mpc}
    \\
    &e_i^{\;T} x_{i,n} \geq e_{del\_min},
        \label{eqn:deliveredmin mpc actual}
        \;\;\forall i=1,\dots,I,
    \\
        &\nonumber\hspace{19.5ex}\forall n=1,\dots,N,
    \\
        &e_j^{\;T} x_{j} \geq e_{del\_min},
        \label{eqn:deliveredmin mpc model}
        \hspace{1.5ex}\forall j=I+1,\dots,I+J,
    \\
    &\sum_{k=1}^{I+J} e_k(t) \leq e_{trans}, \;\;\hspace{20pt} \forall t=1,\dots,T,
    \label{eqn:transformer}
    \\
    &\hat{e}_{inc}\geq \sum_{k=1}^{I+J} e_k(t) - \hat{e}_{old}, \hspace{7pt}\forall t=1,\dots,T,
    \label{eqn: peak 1 }
    \\
    &\hat{e}_{inc}\geq0.
    \label{eqn: peak 2}
\end{align}
\end{subequations}

The first term of the objective function \eqref{eqn:obj mpc} accounts for all $I$ EVs currently plugged in and their $N$ potential departure times each while the second term of the objective function accounts for all $J$ EVs in the future model. The third term ($w_2 p_d \hat{e}_{inc}$) accounts for the demand charge from increasing the current month's peak demand ($\hat{e}_{old}$) by $\hat{e}_{inc}$. Constraint \eqref{eqn:0 mpc} ensures that the energy delivered is nonnegative and less than  the EVSE max $e_{max}$. Constraint \eqref{eqn:deliveredmin mpc actual} ensures a minimum amount of energy is delivered to each EV currently plugged in, constraint \eqref{eqn:deliveredmin mpc model} ensures a minimum amount of energy is delivered to each EV in the future model, and constraint \eqref{eqn:transformer} ensures that the sum of all energy delivered by the EVSEs at each time $t$ does not exceed the transformer constraint $e_{trans}$. Constraints \eqref{eqn: peak 1 }-\eqref{eqn: peak 2} keep track of any increase to the current month's peak load for the demand charge. We note that $\hat{e}_{old}$ corresponds to the previous peak energy demand that has been observed during the month. The pseudocode for the daily algorithm can be viewed in Algorithm \ref{algorithm} below.

\begin{algorithm}[]
\begin{footnotesize}
    \caption{\textsc{Real-Time Smart Charging}}
    \label{algorithm}
    \begin{algorithmic}[1]
    \FOR{ each day}
            \STATE Update current parking lot state 
        \FOR{ each 15 minute interval $t$}
            \IF{new departure from parking lot}
                \STATE Update parking lot state
            \ENDIF
            \IF{new arrival to parking lot}
                \STATE Generate $N$ potential departure times for new arrival
                \STATE Update Parking lot state
            \ENDIF
            \STATE \textbf{Formulate optimization for time $t$:}
            \FOR{each EV $i$ plugged in at time $t$}
                \STATE Add EV $i$ to total objective function 
                \eqref{eqn:obj mpc}
                \STATE Add EV $i$ to active constraints \eqref{eqn:0 mpc}-\eqref{eqn: peak 2}
            \ENDFOR
            \FOR{each future EV $j$ in daily model $t_{model}>t$}
                \STATE Add EV $j$ to total objective function \eqref{eqn:obj mpc}
                \STATE Add EV $j$ to active constraints \eqref{eqn:0 mpc}-\eqref{eqn: peak 2}
            \ENDFOR
            \STATE \textbf{Solve optimization \eqref{eqn:obj mpc}-\eqref{eqn: peak 2} for time $t$}
            \STATE Store planned energy schedule for each EV $i$
            \STATE Set each EVSE's output power for the current 15 minute interval
            \STATE Update peak load $\hat{e}_{old}$ for demand charge calculation (if a new peak load is observed)
        \ENDFOR
    \ENDFOR
    \end{algorithmic}
\end{footnotesize}
\end{algorithm}

\section{Case Study Results}
\label{section: case study}

We examine a two week period from June 17 - June 29 in 2019 at a Bay Area workplace from our Google EV dataset. The location has 57 level 2 EVSEs  \textcolor{black}{with 50-100 EVs arriving each weekday} and is under PG\&E's E-19 rate structure. 

\begin{figure}[]
    \centering
    \includegraphics[width=0.85\columnwidth]{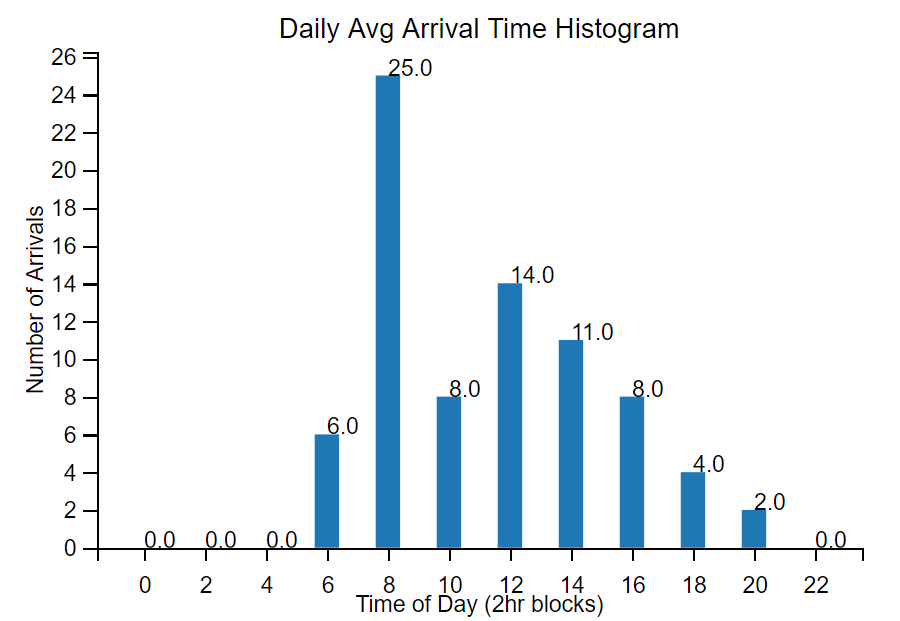}
    \caption{Average arrival time histogram for weekdays in 2019, split into 2 hour blocks, to be used for the daily model. }
    \label{fig: hist}
\end{figure}

First, the EV charging session data was filtered by weekday and then filtered again by arrival time. Namely, each charging session was put into one of 12 possible groups corresponding to 2-hour windows for the arrival times (e.g., an EV charging session that started at 9:48am would be stored in the 8:00am-10:00am group). Once this was done, daily arrival time histograms were generated and the average stay duration and average energy consumption were calculated for each of the 12 groups. The average arrivals per weekday, the average arrivals per 2 hour window, the groups' average stay durations, and the groups' average energy consumption were then used to create the algorithm's future model each day and to generate potential departure times for each EV arrival. The daily average arrival time histogram can been viewed in Figure \ref{fig: hist}. The 2 hour groups' average energy consumption and average stay duration can be viewed in Table I. Last, we note that all of these simulations were done in Python with CVX and Mosek on a Laptop with an i7 processor and 16gb of RAM. Solving the daily smart charging optimization \eqref{eqn:obj mpc}-\eqref{eqn: peak 2} at each 15 minute time step took less than a second, so the algorithm is fit to run in real-time. \textcolor{black}{Moreover, the optimization problem’s complexity is not affected by the number of arriving EVs each day; rather, the problem size grows only as the number of chargers increases. Additionally, for implementation, (2a)-(2g) has to be solved every 15 minutes but for the 57 chargers in our case study, (2a)-(2g) was solved in less than a second. Thus, the algorithm is scalable and there is significant extra time for computation for a larger dataset (i.e., more chargers at the parking lot).}



\begin{table}[h!]
\begin{center}
\resizebox{0.85\columnwidth}{!}{%
\begin{tabular}{||l c c ||} 
 \hline
 Arrival Time Block & Avg Charge Amount (kWh) & Avg Stay Duration \\ [0.5ex] 
 \hline\hline
 12:00am-2:00am & 5.56 & 4 hrs 38 mins  \\ 
 \hline
 2:00am-4:00am & 4.00 & 2 hrs 02 mins  \\ 
 \hline
 4:00am-6:00am & 12.91 & 3 hrs 52 mins  \\ 
 \hline
 6:00am-8:00am & 14.63 & 5 hrs 29 mins  \\ 
 \hline
 8:00am-10:00am & 15.79 & 6 hrs 02 mins  \\ 
 \hline
 10:00am-12:00pm & 9.27 & 6 hrs 02 mins  \\ 
 \hline
 12:00pm-2:00pm & 7.41 & 11 hrs 15 mins  \\ 
 \hline
 2:00pm-4:00pm & 6.80 & 16 hrs 06 mins  \\ 
 \hline
 4:00pm-6:00pm & 7.14 & 16 hrs 27 mins  \\ 
 \hline
 6:00pm-8:00pm & 6.61 & 23 hrs 19 mins  \\ 
 \hline
 8:00pm-10:00pm & 6.78 & 25 hrs 54 mins  \\ 
 \hline
 10:00pm-12:00am & 7.74 & 10 hrs 01 mins  \\ 
 \hline
 \hline
\end{tabular}}
\end{center}
\caption{\textrm{Average charge amount (kwh) and average stay duration for different arrival time blocks.}}
\end{table}

\subsection{Example Day's Charging Schedule Evolution}
\label{sec: evolution}

\begin{figure}[]
    \centering
    \includegraphics[width=0.97\columnwidth]{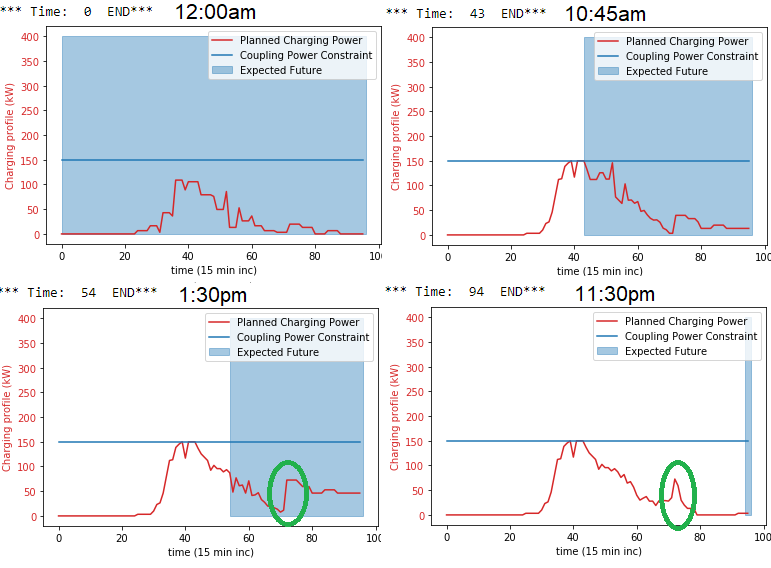}
    \caption{Evolution of planned charging power. Top Left: At 12:00am the algorithm is fully using the expected future as a model for what could happen that day. Top Right: At 10:45am, many EVs have arrived and the power output has started to deviate from the expected future. Bottom Left: Planned power output at 1:30pm. Bottom Right: At 11:30pm, there are no more arrivals, Red curve shows the daily power output of the 57 EVSEs. Additionally, the green circles indicate spikes in total charging power in the evening due to the algorithm waiting for cheap electricity rates to charge multiple EVs, discussed in Section \ref{sec: infeasibilities}.}
    \label{fig: intuitive}
\end{figure}

Figure \ref{fig: intuitive} shows an example as to how the real-time optimization changes the planned power output as time progresses and more EVs arrive. These plots correspond to a Monday with 78 EVs arriving to 57 EVSEs with $(w_1, w_2)=(2,1)$ and a transformer capacity constraint of 150kW. Starting with the Top Left: This plot shows the planned charging schedule that is calculated at 12:00am on the given Monday for the entire location. There are no actual EV arrivals at this early time in the morning so the entire planned charging schedule is created by looking at the future model for the day (shaded blue region). Top Right: This shows the planned charging power as of 10:45am in the morning. Everything that is not in the shaded blue region corresponds to time-steps in the past, meaning that the red plot in the white region corresponds to actual charging power. However, everything in the blue region is still estimated via the future model. Bottom Left: This plot shows the planned charging schedule as of 1:30pm. Bottom Right Plot: This plot shows the charging schedule as of 11:30pm, when there are no more arrivals for that day. Additionally, the smart charging algorithm is able to ensure that the actual charging schedule stays below the transformer capacity constraint.

\begin{figure}[]
    \centering
    \includegraphics[width=0.98\columnwidth]{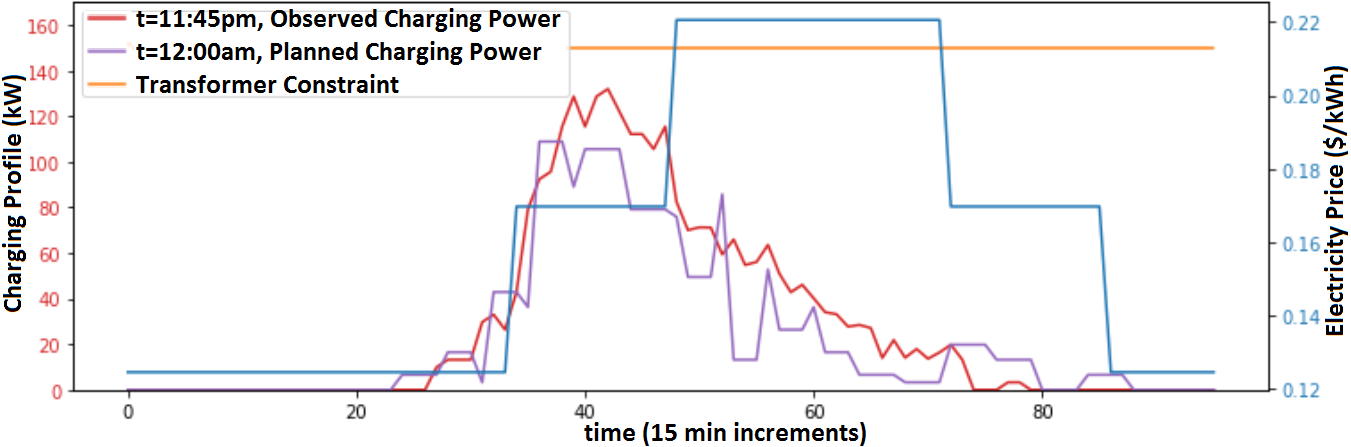}
    \caption{A comparison of the model's predicted daily charging schedule vs. the actual daily charging power.}
    \label{fig: model comparison}
\end{figure}

Figure \ref{fig: model comparison} presents a comparison of a weekday's predicted daily charging schedule and the actual daily charging power that ocurred that day. As seen in Figure \ref{fig: model comparison} the daily model does a good job predicting the future load for this location. Furthermore, note that most of the power is scheduled to be delivered during the partial-peak electricity rates during the mid-morning (8:30am-12:00pm) and the charging power decreases rapidly during the peak electricity rates mid-day (12:00pm-6:00pm).

\subsection{Test Case Comparisons}
\label{sec: comparisons}

Furthermore, in Table II, we present simulation results for several test cases. In these tests, we vary $(w_1,w_2)$ in the objective function between (2,1) and (10,1) to show how the algorithm swaps priority from minimizing electricity cost to maximizing user utility for receiving energy. Additionally, we varied the transformer constraint that couples all the EVs together between 250kW and 100kW. In Table II, we also show the total energy delivered over the 2 week period and the total cost of purchasing that energy from the grid from energy rates as well as the total cost due to demand charges (displayed as percentages compared to the status quo values). The last 2 columns in Table II show whether or not the test case included a constraint that forces the EVSEs to charge at a certain rate for the first hour that a new EV is plugged in. The idea behind this is to ensure that EVs will receive some minimum amount of energy, even if they are only plugged in for a short duration, or if they arrive during peak electricity rates. We experimented with forcing the EVs to charge for the first hour of a new session at their maximum rate, half their maximum rate, a quarter of their maximum rate, and without this constraint altogether. The second to last column indicates whether or not this constraint was included, and the last column indicates the charge rate that was used for this first hour ($p_{max}$ = 6.6kW). The first row of this table shows the total energy delivered and the total energy cost for the status quo (i.e., no algorithm in place, just the energy consumed and cost for June 17-June 29, 2019). The $(w_1, w_2)=(10,1)$ results in Tests 07-13 indicate that the smart charging algorithm was able to reduce energy costs and demand charges while delivering adequate energy (greater than $80\%$ of the status quo energy) to the EVs.


\setlength{\tabcolsep}{3pt}
\begin{table}[]
\begin{center}
\resizebox{0.97\columnwidth}{!}{%
\begin{tabular}{||c c c c c c c c c ||} 
 \hline
 Test \# & $w_1$ & $w_2$ & \pbox{20cm}{Coupling \\ Constraint \\(kW)} & \pbox{20cm}{Energy \\ Delivered} & \pbox{20cm}{Electricity \\ Purchase Cost} & \pbox{20cm}{Demand\\Charge Cost}& \pbox{20cm}{Forced \\Initial \\Charge} & \pbox{20cm}{Initial \\Charge \\Rate}\\ [0.1ex] 
 \hline\hline
  \pbox{20cm}{Status Quo} & n/a & n/a & n/a & 100\% & 100\% & 100\% & n/a & n/a \\ 
 \hline
 01 & 2 & 1 & 250 & 50.13\% & 45.78\% & 64.50\% & Yes & $\frac{1}{2}p_{max}$ \\ 
 \hline
  02 & 2 & 1 & 150 & 50.01\% & 45.83\% & 88.76\%& Yes & $\frac{1}{2}p_{max}$ \\ 
 \hline
  03 & 2 & 1 & 125 & 50.68\% & 46.33\% &73.96\%& Yes & $\frac{1}{2}p_{max}$ \\ 
 \hline
  04 & 2 & 1 & 110 & 50.44\% & 45.94\% &65.09\%& Yes & $\frac{1}{2}p_{max}$ \\ 
 \hline
  05 & 2 & 1 & 110 & 44.20\% & 36.40\% &65.09\%& No & n/a \\ 
 \hline
  06 & 2 & 1 & 100 & 45.05\% & 37.45\% &59.17\%& No & n/a \\ 
 \hline
   07 & 10 & 1 & 100 & 81.93\% & 81.36\% &59.17\%& No & n/a \\ 
 \hline
    08 & 10 & 1 & 150 & 83.85\% & 81.03\% & 88.76\%& No & n/a \\ 
 \hline
     09 & 10 & 1 & 250 & 84.21\% & 80.98\% &89.94\%& No & n/a \\ 
 \hline
     10 & 10 & 1 & 100 & Infeasible & Infeasible &Infeasible& Yes & $\frac{1}{2}p_{max}$ \\ 
 \hline
     11 & 10 & 1 & 110 & 81.76\% & 81.09\% &65.09\%& Yes & $\frac{1}{2}p_{max}$ \\
\hline
12 & 10 & 1 & 150 & 83.49\% & 81.80\% & 88.76\%& Yes & $\frac{1}{2}p_{max}$ \\
 \hline 
13 & 10 & 1 & 150 & Infeasible & Infeasible & Infeasible& Yes & $p_{max}$ \\
 \hline 
13 & 10 & 1 & 250 & 87.25\% & 84.81\% &97.63\%& Yes & $p_{max}$ \\
 \hline
\end{tabular}}
\end{center}
\caption{Results for 13 different test cases.}
\end{table}

\subsection{Transformer Capacity Constraints}
\label{sec: transformer}

\begin{figure}[]
    \centering
    \includegraphics[width=0.94\columnwidth]{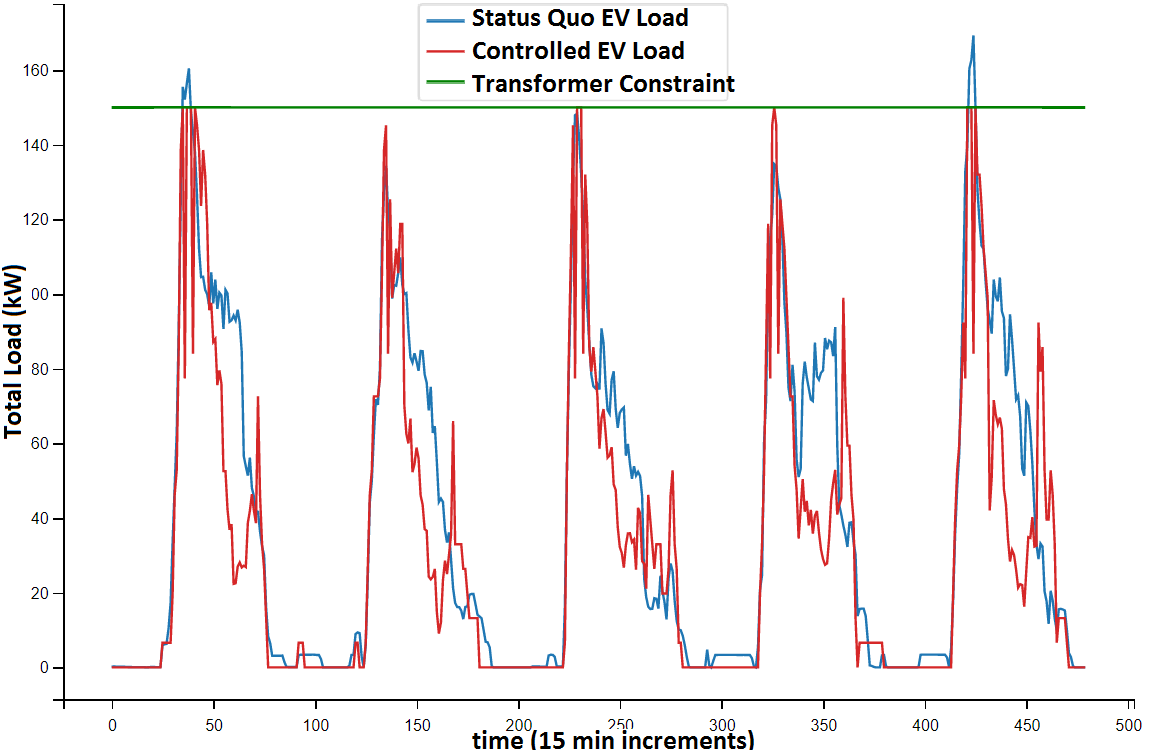}
    \caption{Comparison of the Monday-Friday weekday load for the status quo (i.e., no algorithm in place) vs. the smart charging algorithm with parameters as listed in Test 08.  Additionally, the smart charging algorithm enforces the transformer constraint (green line).}
    \label{fig: 5 day}
\end{figure}

As shown in Table II, we vary the transformer capacity constraint that couples the charging power of all the EVs. In the status quo, there is no coupling constraint and the total load peaks at 169 kW. However, with a smart charging algorithm in place, we can constrain the total load. This would allow for the location to use a smaller transformer capacity or increase their other non-EVs loads (e.g., the nearby offices  can safely use more power without worrying about exceeding the transformer capacity due to the EVSEs). Figure \ref{fig: 5 day} shows a 5 day comparison of the Test 08 load (see Table II) compared to the status quo load. As seen in this plot, the smart charging algorithm was able to enforce a transformer capacity limit at 150kW without sacrificing much energy delivered. Furthermore, the red curve (the smart charging profile) drops below the blue curve (the status quo) during the middle of each day to avoid the peak electricity rates from 12:00pm-6:00pm.

\subsection{Infeasibilities Due to Transformer Capacities}
\label{sec: infeasibilities}

Additionally, we would like to discuss the effects of removing the constraint forcing the EVSEs to charge at a certain rate for the first hour an EV is plugged in. Recall, this constraint was added to ensure that EVs are charged before they depart. If this constraint is removed, the smart charging algorithm becomes overly optimistic about each EV's departure time. Specifically, the algorithm optimistically predicts that the EVs will stay until the off-peak electricity rates in the evening and plans the daily charging schedule as seen in Plots 3,4 of Figure \ref{fig: intuitive} (circled in green). Including the constraint that forces the EVSEs to charge a new arrival for the first hour removes the second peak that is seen in Plots 3,4 of Figure \ref{fig: intuitive}.

\begin{figure}[]
    \centering
    \includegraphics[width=0.96\columnwidth]{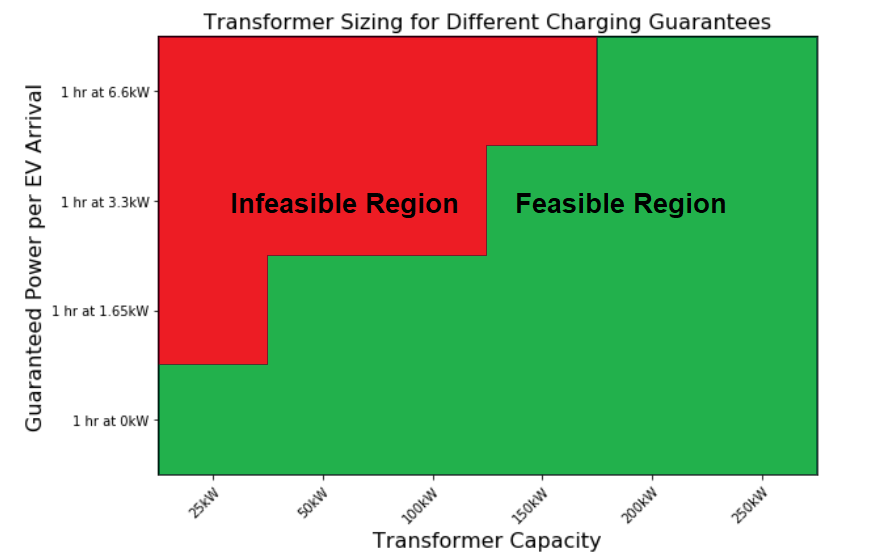}
    \caption{The feasible and infeasible regions of operation for various transformer capacities and forced initial charge rates for new EV arrivals.}
    \label{fig: regions}
\end{figure}

The constraint that forces EVSEs to charge new EV arrivals for the first hour of being plugged in also affects whether or not the optimization is feasible each day. Namely, many EVs arrive around the start of the workday at 8:00am and plug in to an EVSE. If many EVSEs are forced to charge near their maximum rating at the same time, the total load of the parking lot can exceed the transformer capacity. As such, in Figure \ref{fig: regions} we show the feasible and infeasible regions of operation for the smart charging algorithm with $w_1$=10 and $w_2$=1.

\section{Conclusion}
In this paper, we presented a smart charging algorithm for a workplace parking lot equipped with EVSEs that operates in real-time to minimize electricity cost from time-of-use electricity rates and demand charges while ensuring that the owners of the EVs receive adequate levels of charge. Our algorithm uses both scenario generation to account for each EV's unknown future departure time as well as certainty equivalent model predictive control to account for the unknown EV arrivals in the future. We build models from our Google dataset for each day of the week and our algorithm uses these models as the expected future when optimizing the EVs charging schedules. 
In the future, we have access to the meter data for loads other than the EVSEs for all locations in our dataset as well as the single-line diagram specifications and we hope to include more operational constraints and account for the non-EV loads in the smart charging algorithm. Furthermore, some locations have  energy storage and solar generation which we hope to include into future algorithms. Finally, future work will deploy this framework in real-time field operation.


\begin{small}
    \textsc{Acknowledgment:} This work was funded by the California Energy Commission under grant EPC-17-020. SLAC National Accelerator Laboratory is operated for the US Department of Energy by Stanford University  under  Contract  DE-AC02-76SF00515.  Thank you to Rolf Schreiber from Google for providing data.    
\end{small}



%

\linespread{0.9}
\bibliographystyle{IEEEtran}
\bibliography{references}

\end{document}